\documentclass[showpacs,preprintnumbers,twocolumn,superscriptaddress,aps,nofootinbib]{revtex4-2}
\usepackage{amssymb}
\usepackage[centertags]{amsmath}
\usepackage{graphicx,subfigure}
\usepackage{txfonts}
\usepackage{epsfig}
\usepackage{bm}
\usepackage{color}
\usepackage{graphicx,graphics}
\usepackage{multirow}
\usepackage{float}
\usepackage{ulem}
\usepackage{appendix}
\usepackage[pdfstartview=FitH]{hyperref}
\hypersetup{colorlinks=true, citecolor=blue,linkcolor=blue,filecolor=black,urlcolor=blue}  
\allowdisplaybreaks[2]

\begin{document}

\title{Production characteristics of light (anti-)nuclei from (anti-)nucleon coalescence in heavy ion collisions at energies employed at the RHIC beam energy scan}

\author{Xiang-Yu Zhao}
\affiliation{School of Physics and Physical Engineering, Qufu Normal University, Shandong 273165, China}

\author{Yan-Ting Feng}
\affiliation{School of Physics and Physical Engineering, Qufu Normal University, Shandong 273165, China}

\author{Feng-Lan Shao}
\email {shaofl@mail.sdu.edu.cn}
\affiliation{School of Physics and Physical Engineering, Qufu Normal University, Shandong 273165, China}

\author{Rui-Qin Wang}
\email {wangrq@qfnu.edu.cn}
\affiliation{School of Physics and Physical Engineering, Qufu Normal University, Shandong 273165, China}

\author{Jun Song}
\email {songjun2011@jnxy.edu.cn}
\affiliation{School of Physical Science and Intelligent Engineering, Jining University, Shandong 273155, China}

\begin{abstract}

With the kinetic freeze-out nucleons and antinucleons obtained from the quark combination model, we study the production of light nuclei and antinuclei in the (anti-)nucleon coalescence mechanism in relativistic heavy ion collisions.
We derive analytic formulas of the momentum distributions of different light nuclei and apply them to compute transverse momentum ($p_T$) spectra of (anti-)deuterons ($d$, $\bar d$) and (anti-)tritons ($t$, $\bar t$) in Au-Au collisions at $\sqrt{s_{NN}}=$7.7, 11.5, 19.6, 27, 39, 54.4 GeV.
We find that the experimental data available for these $p_T$ spectra can be well reproduced.
We further study the yields and yield ratios of different light (anti-)nuclei and naturally explain their interesting behaviors as a function of the collision energy.
We especially point out that the multi-particle yield ratio $tp/d^2$ should be carefully corrected from hyperon weak decays for protons to probe the production characteristics of light nuclei.
All of our results show that the coalescence mechanism for (anti-)nucleons plays a dominant role for the production of light nuclei and antinuclei at the RHIC beam energy scan energies.

\end{abstract}

\maketitle

\section{Introduction}

Light nuclei and antinuclei such as (anti-)deuterons and (anti-)tritons are considered to be a unique kind of probes in ultra-relativistic heavy ion collisions.
On one hand, they can effectively explore the information of the bulk system, especially the system freeze-out properties such as the geometrical freeze-out volume~\cite{1BA1994PRL,YGMa2018PRept} and freeze-out particle correlations~\cite{6BA2019PRC}, etc., since they are mostly produced at the late stage of the system evolution.
On the other hand, the production of such composite particles itself is very much well worth studying and it is closely related with many fundamental issues in high energy physics and in astronomy field, e.g., the hadronization mechanism~\cite{Aichelin1991PRept}, cosmic-ray production and propagation in the Galaxy~\cite{Cosmicray2020JCAP}, etc.

In recent years the theoretical study of the production of light (anti-)nuclei has re-absorbed much attention in heavy ion collisions ~\cite{QCDphaseKLS2018PLB,QGP2019CERNYellow,QCDphaseNuXu2020PRept,review2019NPA,Oliinychenko2021NPA}.
Two production mechanisms have proved to be particularly successful in describing the light nuclei formation.
One is the thermal production~\cite{1thermal1977PRL,2thermal1979PRL,3thermal2011PLB,4thermal2011PRC,5thermal2018nature,Tthermal2019PRC},
which assumes that (anti-)nuclei are produced from a thermally and chemically equilibrated source like abundantly-produced mesons and baryons.
The other is the coalescence mechanism~\cite{1coale1963PR,2coale1963PR,3coale1981PLB,4coale1991PRC,5coale1995PRL,6coale1996PRC,7coale1997PRC,8coale2003PRC,9WBZhao2018PRC,PRC1980,1BA1994PRL,2BA1998PLB,3BA1999PRC,4BA2018PRC,5BA2018MPLA,6BA2019PRC,finalrecom1976PRL},
in which light (anti-)nuclei are assumed to be produced by the coalescence of the jacent (anti-)nucleons in the phase space.
Such production mechanism possesses its unique characteristics.
Many specific models and/or event-generators such as hybrid dynamical model (iEBE-MUSIC)~\cite{ZhaoWB2020PRC},  the Ultra-relativistic-Quantum-Molecular-Dynamics model (UrQMD)~\cite{UrQMD2020PLB},  Jet AA Mi-croscopic Transportation Model (JAM)~\cite{PLB805LiuHui}, the parton and hadron cascade model (PACIAE) ~\cite{PACIAE2019EPJA}, etc., have been developed to include light nuclei formation via the nucleon coalescence and provided nice explanations for series of observables.
Besides these two mechanisms, transport scenario is also proposed for light nuclei production, which assumes the existence of light nuclei in strongly-interacting hadronic matter and aims to study how light nuclei evolve during the hadronic system evolution ~\cite{another2009PRC,another2019PRC,another2021PRC1,another2021PRC2}.

Experiments at the BNL Relativistic Heavy Ion Collider (RHIC) and the CERN Large Hadron Collider (LHC) have accumulated a wealth of data on light nuclei production.
These data exhibit some fascinating features~\cite{v2phid2007PRLPHENIX,v2d2016PRCSTAR,v1d2020PRCSTAR,DWZhang2021NPASTAR,B2B32001PRLSTAR,B2d2019PRCSTAR}, especially their non-trivial energy-dependent behaviors at RHIC energies. 
Such behaviors are considered to be possible signals for the critical end point (CEP) of the first order phase transition from hadronic phase to quark-gluon phase in some works such as in Refs.~\cite{PLB805LiuHui,KJS2021EPJA}.
As we know, the whole process of the relativistic heavy-ion collision is a very complicated process, involving many components, e.g., hard parton scatterings, collective expansion evolution, hadronization, hadronic rescatterings, resonance decays and so on.
These components are very different in different centrality collisions at different energies and they finally lead to very different hadronic systems at the kinetic freeze-out.
Is the nucleon coalescence a universal mechanism for light nuclei production in these different hadronic systems? 
Whether the above non-monotonic energy-dependent behaviors of light nuclei, usually taken as possible signals for the CEP, 
are caused by the differences of the hadronic systems at different collision energies or by the production mechanism itself?

In this article, we apply the coalescence mechanism to hadronic systems created in Au-Au collisions at RHIC energies to study the production of light (anti-)nuclei in the low- and intermediate-$p_T$ regions. 
One main goal of this article is to bring to light the characteristics originating mainly from the nucleon coalescence and to discriminate influences of different factors in heavy ion collisions on light nuclei production.
For this purpose, we begin with the kinetic freeze-out nucleons and antinucleons obtained from the quark combination model developed by the Shandong Group ~\cite{Song2021PRC,yanting work}, and then we let these nucleons coalescence into different (anti-)nuclei to study their production characteristics.
We find that weak decay contaminations for protons from $\Lambda$ and $\Xi$ hyperons are different in different centralities at different collision energies,
and this should be carefully considered when using some light nuclei yield ratios related with protons measured by the STAR Collaboration such as the multi-particle yield ratio $tp/d^2$ to probe the production characteristics of light nuclei and extract CEP signal.

The rest of the article is organized as follows. 
In Sec.~II, we give an introduction to the derivation of the momentum distributions of light nuclei in the framework of the nucleon coalescence. 
In Sec.~III, we systematically study the $p_T$ spectra and midrapidity yield densities of  $d$, $\bar d$, $t$, $\bar t$ in different centralities in Au-Au collisions at $\sqrt{s_{NN}}=$7.7, 11.5, 19.6, 27, 39, 54.4 GeV.
We present in particular various yield ratios of light nuclei such as $d/p$, $\bar d/\bar p$, $t/p$, $\bar t/\bar p$, $d/p^2$, $\bar d/\bar p^2$, $tp/d^2$, etc. and discuss their properties as functions of the collision energy and the collision centrality. 
In Sec.~IV, we give our summary. 

%
\section{The nucleon coalescence model}   \label{model}

In this section we briefly introduce the nucleon coalescence model, which is used to deal with the formation of light (anti-) nuclei.
This model has been successfully used to explain nontrivial behaviors of the coalescence factor measured in different collision systems at the CERN Large Hadron Collider~\cite{RQWang2021PRC}.

We start from a hadronic system produced at the final stage of the evolution of high energy collision and suppose light nuclei are formed via the nucleon coalescence.
The three-dimensional momentum distribution of the produced deuterons $f_{d}(\bm{p})$ and that of tritons $f_{t}(\bm{p})$ are given by
{\setlength\arraycolsep{0pt}
\begin{eqnarray}
 f_{d}(\bm{p})&=& N_{pn} \int d\bm{x}_1d\bm{x}_2 d\bm{p}_1 d\bm{p}_2  f^{(n)}_{pn}(\bm{x}_1,\bm{x}_2;\bm{p}_1,\bm{p}_2)  \nonumber  \\
   &&~~~~~~~~~ \times  \mathcal {R}_{d}(\bm{x}_1,\bm{x}_2;\bm{p}_1,\bm{p}_2,\bm{p}),      \label{eq:fdgeneral}   \\
 f_{t}(\bm{p})&=& N_{pnn} \int d\bm{x}_1d\bm{x}_2d\bm{x}_3 d\bm{p}_1 d\bm{p}_2 d\bm{p}_3 \nonumber  \\
   &&~~~~~~~~~~ \times  f^{(n)}_{pnn}(\bm{x}_1,\bm{x}_2,\bm{x}_3;\bm{p}_1,\bm{p}_2,\bm{p}_3) \nonumber  \\
   &&~~~~~~~~~~ \times \mathcal {R}_{t}(\bm{x}_1,\bm{x}_2,\bm{x}_3;\bm{p}_1,\bm{p}_2,\bm{p}_3,\bm{p}),      \label{eq:ftgeneral} 
\end{eqnarray} }%
where $f^{(n)}_{pn}(\bm{x}_1,\bm{x}_2;\bm{p}_1,\bm{p}_2)$ and $f^{(n)}_{pnn}(\bm{x}_1,\bm{x}_2,\bm{x}_3;\bm{p}_1,\bm{p}_2,\bm{p}_3)$ are normalized two- and three- nucleon joint coordinate-momentum distributions, respectively; 
$N_{pn}=N_{p}N_{n}$ is the number of all possible $pn$-pairs and $N_{pnn}=N_{p}N_{n}(N_{n}-1)$ is that of all possible $pnn$-clusters;
$N_{p}$ is the number of protons and $N_{n}$ is that of neutrons in the considered hadronic system.
$\mathcal {R}_{d}(\bm{x}_1,\bm{x}_2;\bm{p}_1,\bm{p}_2,\bm{p})$ and $\mathcal {R}_{t}(\bm{x}_1,\bm{x}_2,\bm{x}_3;\bm{p}_1,\bm{p}_2,\bm{p}_3,\bm{p})$ are kernel functions.
Here and from now on we use bold symbols to denote three-dimensional coordinates and momenta.

Kernel functions $\mathcal {R}_{d}(\bm{x}_1,\bm{x}_2;\bm{p}_1,\bm{p}_2,\bm{p})$ and $\mathcal {R}_{t}(\bm{x}_1,\bm{x}_2,\bm{x}_3;\bm{p}_1,\bm{p}_2,\bm{p}_3,\bm{p})$ 
carry the kinetic and dynamical information of the nucleons recombining into light nuclei,
and their precise expressions should be constrained by such as the momentum conservation, constraints due to intrinsic quantum numbers e.g. spin, and so on~\cite{RQWang2019CPC}.
To take these constraints into account explicitly, we rewrite them in the following forms
{\setlength\arraycolsep{0pt}
\begin{eqnarray}
&&  \mathcal {R}_{d}(\bm{x}_1,\bm{x}_2;\bm{p}_1,\bm{p}_2,\bm{p}) = g_d \mathcal {R}_{d}^{(x,p)}(\bm{x}_1,\bm{x}_2;\bm{p}_1,\bm{p}_2) \delta(\displaystyle{\sum^2_{i=1}} \bm{p}_i-\bm{p}),  \nonumber  \\ \label{eq:Rd}  \\
&&  \mathcal {R}_{t}(\bm{x}_1,\bm{x}_2,\bm{x}_3;\bm{p}_1,\bm{p}_2,\bm{p}_3,\bm{p}) = g_{t}  \nonumber  \\
&&~~~~~~~~~~~~~~~~ \times \mathcal {R}_{t}^{(x,p)}(\bm{x}_1,\bm{x}_2,\bm{x}_3;\bm{p}_1,\bm{p}_2,\bm{p}_3)   \delta(\displaystyle{\sum^3_{i=1}} \bm{p}_i-\bm{p}) ,      \label{eq:Rt}  
\end{eqnarray} }%
where the spin degeneracy factors $g_d=3/4$ and $g_{t}=1/4$.
The Dirac $\delta$ functions guarantee the momentum conservation in the coalescence.
The remaining $\mathcal {R}_{d}^{(x,p)}(\bm{x}_1,\bm{x}_2;\bm{p}_1,\bm{p}_2)$ can be solved from the Wigner transformation once the wave functions of the light nuclei are given with the instantaneous coalescence approximation.
They are as follows
{\setlength\arraycolsep{0pt}
\begin{eqnarray}
&&  \mathcal {R}^{(x,p)}_{d}(\bm{x}_1,\bm{x}_2;\bm{p}_1,\bm{p}_2) = 8e^{-\frac{(\bm{x}'_1-\bm{x}'_2)^2}{\sigma_d^2}}     e^{-\frac{\sigma_d^2(\bm{p}'_{1}-\bm{p}'_{2})^2}{4\hbar^2c^2}},      \label{eq:Rdxp}  \\
&&  \mathcal {R}^{(x,p)}_{t}(\bm{x}_1,\bm{x}_2,\bm{x}_3;\bm{p}_1,\bm{p}_2,\bm{p}_3) =8^2e^{-\frac{(\bm{x}'_1-\bm{x}'_2)^2}{2\sigma_{t}^2}} e^{-\frac{(\bm{x}'_1+\bm{x}'_2-2\bm{x}'_3)^2}{6\sigma_{t}^2}}  \nonumber   \\
&& ~~~~~~~~~~~~~~~~~~~~~~~~~~\times e^{-\frac{\sigma_{t}^2(\bm{p}'_{1}-\bm{p}'_{2})^2}{2\hbar^2c^2}} e^{-\frac{\sigma_{t}^2(\bm{p}'_{1}+\bm{p}'_{2}-2\bm{p}'_{3})^2}{6\hbar^2c^2}},  \label{eq:Rtxp}
\end{eqnarray} }%
as we adopt the wave function of a spherical harmonic oscillator as in Refs.~\cite{Wigner2003NPA,Wigner2015PRC}.
The superscript `$'$' in the coordinate or momentum variable denotes the coordinate or momentum of the nucleon in the rest frame of the $pn$-pair or $pnn$-cluster.
The width parameter $\sigma_d=\sqrt{\frac{8}{3}} R_d$ and $\sigma_{t}=R_{t}$,
where $R_d$ and $R_{t}$ are the root-mean-square radius of the deuteron and that of the triton, respectively.
The factor $\hbar c$ comes from the used GeV$~$fm unit, and it is 0.197 GeV$~$fm.

For the normalized joint distributions of the nucleons, we consider a simple case that they are coordinate and momentum factorized, i.e.,
{\setlength\arraycolsep{0pt}
\begin{eqnarray}
&& f^{(n)}_{pn}(\bm{x}_1,\bm{x}_2;\bm{p}_1,\bm{p}_2) = f^{(n)}_{pn}(\bm{x}_1,\bm{x}_2)   f^{(n)}_{pn}(\bm{p}_1,\bm{p}_2),  \label{eq:fpnfac}  \\
&& f^{(n)}_{pnn}(\bm{x}_1,\bm{x}_2,\bm{x}_3;\bm{p}_1,\bm{p}_2,\bm{p}_3) = f^{(n)}_{pnn}(\bm{x}_1,\bm{x}_2,\bm{x}_3)  \nonumber  \\
&&~~~~~~~~~~~~~~~~~~~~~~~~~~~~~~~ \times  f^{(n)}_{pnn}(\bm{p}_1,\bm{p}_2,\bm{p}_3).  \label{eq:fpnnfac}    
\end{eqnarray} }%
Substituting Eqs.~(\ref{eq:Rd}-\ref{eq:fpnnfac}) into Eqs.~(\ref{eq:fdgeneral}) and (\ref{eq:ftgeneral}), we have
{\setlength\arraycolsep{0.2pt}
\begin{eqnarray}
&& f_{d}(\bm{p})= g_{d} N_{pn} \int d\bm{x}_1d\bm{x}_2 f^{(n)}_{pn}(\bm{x}_1,\bm{x}_2) 8e^{-\frac{(\bm{x}'_1-\bm{x}'_2)^2}{\sigma_d^2}}  \nonumber   \\
&& ~~~ \times
 \int d\bm{p}_1d\bm{p}_2 f^{(n)}_{pn}(\bm{p}_1,\bm{p}_2) e^{-\frac{\sigma_d^2(\bm{p}'_{1}-\bm{p}'_{2})^2}{4\hbar^2c^2}} \delta(\displaystyle{\sum^2_{i=1}} \bm{p}_i-\bm{p}),~~~~~~     \label{eq:fd}  \\
&& f_{t}(\bm{p}) = g_{t} N_{pnn}  \nonumber   \\
&& ~~~ \times \int d\bm{x}_1d\bm{x}_2d\bm{x}_3 f^{(n)}_{pnn}(\bm{x}_1,\bm{x}_2,\bm{x}_3)  
8^2e^{-\frac{(\bm{x}'_1-\bm{x}'_2)^2}{2\sigma_{t}^2}} e^{-\frac{(\bm{x}'_1+\bm{x}'_2-2\bm{x}'_3)^2}{6\sigma_{t}^2}}   \nonumber   \\
&& ~~~ \times
 \int d\bm{p}_1d\bm{p}_2d\bm{p}_3 f^{(n)}_{pnn}(\bm{p}_1,\bm{p}_2,\bm{p}_3) 
 e^{-\frac{\sigma_{t}^2(\bm{p}'_{1}-\bm{p}'_{2})^2}{2\hbar^2c^2}} e^{-\frac{\sigma_{t}^2(\bm{p}'_{1}+\bm{p}'_{2}-2\bm{p}'_{3})^2}{6\hbar^2c^2}}   \nonumber \\
&&~~~~~~~~~~~~~~~~~~~~~~~~~~~~~~ \times  \delta(\displaystyle{\sum^3_{i=1}} \bm{p}_i-\bm{p}) .  \label{eq:ft}  
\end{eqnarray} }%
Eqs.~(\ref{eq:fd}) and (\ref{eq:ft}) show that we can calculate momentum distributions of different light nuclei by integrating coordinates and momenta of nucleons, respectively.

We use $\mathcal {A}_d$ and $\mathcal {A}_{t}$ to denote the coordinate integral parts in Eqs.~(\ref{eq:fd}) and (\ref{eq:ft}) as
{\setlength\arraycolsep{0pt}
\begin{eqnarray}
&& \mathcal {A}_d =  8\int d\bm{x}_1d\bm{x}_2 f^{(n)}_{pn}(\bm{x}_1,\bm{x}_2)  e^{-\frac{(\bm{x}'_1-\bm{x}'_2)^2}{\sigma_d^2}},     \label{eq:Ad}  \\
&& \mathcal {A}_{t} = 8^2\int d\bm{x}_1d\bm{x}_2d\bm{x}_3 f^{(n)}_{pnn}(\bm{x}_1,\bm{x}_2,\bm{x}_3)   \nonumber  \\
&& ~~~~~~~~~~~~~~~~~~~~~ \times   e^{-\frac{(\bm{x}'_1-\bm{x}'_2)^2}{2\sigma_{t}^2}} e^{-\frac{(\bm{x}'_1+\bm{x}'_2-2\bm{x}'_3)^2}{6\sigma_{t}^2}}, ~~~~~~  \label{eq:At}       
\end{eqnarray} }%
and use $\mathcal {M}_{d}(\bm{p})$ and $\mathcal {M}_{t}(\bm{p})$ to denote the momentum integral parts as
{\setlength\arraycolsep{0pt}
\begin{eqnarray}
&& \mathcal {M}_d (\bm{p}) =  \int d\bm{p}_1d\bm{p}_2 f^{(n)}_{pn}(\bm{p}_1,\bm{p}_2)  e^{-\frac{\sigma_d^2(\bm{p}'_{1}-\bm{p}'_{2})^2}{4\hbar^2c^2}}   \delta(\displaystyle{\sum^2_{i=1}} \bm{p}_i-\bm{p}),  \nonumber  \\     \label{eq:Md}  \\
&& \mathcal {M}_{t} (\bm{p}) = \int d\bm{p}_1d\bm{p}_2d\bm{p}_3 f^{(n)}_{pnn}(\bm{p}_1,\bm{p}_2,\bm{p}_3)    \nonumber  \\ 
&&~~~~~~~~~~~~~~~~~ \times  e^{-\frac{\sigma_{t}^2(\bm{p}'_{1}-\bm{p}'_{2})^2}{2\hbar^2c^2}} e^{-\frac{\sigma_{t}^2(\bm{p}'_{1}+\bm{p}'_{2}-2\bm{p}'_{3})^2}{6\hbar^2c^2}}    \delta(\displaystyle{\sum^3_{i=1}} \bm{p}_i-\bm{p}).  \label{eq:Mt}       
\end{eqnarray} }%
So we get
{\setlength\arraycolsep{0.2pt}
\begin{eqnarray}
&& f_{d}(\bm{p}) = g_{d} N_{pn} \mathcal {A}_d  \mathcal {M}_{d}(\bm{p}),     \label{eq:fd-AM}  \\
&& f_{t}(\bm{p}) = g_{t} N_{pnn} \mathcal {A}_{t}   \mathcal {M}_{t}(\bm{p}).     \label{eq:ft-AM}  
\end{eqnarray} }%
$\mathcal {A}_d$ stands for the probability of a $pn$- pair satisfying the coordinate requirement to recombine into a deuteron-like molecular state,
and $\mathcal {M}_{d}(\bm{p})$ stands for the probability of a $pn$- pair satisfying the momentum requirement to recombine into a deuteron-like molecular state with momentum $\bm{p}$.
The similar case holds for $\mathcal {A}_{t}$ and $\mathcal {M}_{t}(\bm{p})$.

Changing coordinate integral variables in Eq.~(\ref{eq:Ad}) to be $\bm{X}_C= \frac{\bm{x}_1+\bm{x}_2}{2}$ and $\bm{r}= \bm{x}_1-\bm{x}_2$, 
and those in Eq.~(\ref{eq:At}) to be $\bm{Y}_C= (\bm{x}_1+\bm{x}_2+\bm{x}_3)/\sqrt{3}$, $\bm{r}_1= (\bm{x}_1-\bm{x}_2)/\sqrt{2}$ and $\bm{r}_2= (\bm{x}_1+\bm{x}_2-2\bm{x}_3)/\sqrt{6}$,
we have
{\setlength\arraycolsep{0pt}
\begin{eqnarray}
&& \mathcal {A}_d =  8\int d\bm{X}_C d\bm{r} f^{(n)}_{pn}(\bm{X}_C,\bm{r}) e^{-\frac{\bm{r}'^2}{\sigma_d^2}},   \label{eq:Adr}   \\
&& \mathcal {A}_{t} =  8^2\int d\bm{Y}_Cd\bm{r}_1d\bm{r}_2 f^{(n)}_{ppn}(\bm{Y}_C,\bm{r}_1,\bm{r}_2) 
e^{-\frac{(\bm{r}'_1)^2+(\bm{r}'_2)^2}{\sigma_{t}^2}}. ~~~~  \label{eq:AHer}  
\end{eqnarray} }%
We further assume the coordinate joint distributions are coordinate variable factorized, i.e.,
$f^{(n)}_{pn}(\bm{X}_C,\bm{r}) = f^{(n)}_{pn}(\bm{X}_C) f^{(n)}_{pn}(\bm{r})$ and
$f^{(n)}_{pnn}(\bm{Y}_C,\bm{r}_1,\bm{r}_2) = f^{(n)}_{pnn}(\bm{Y}_C) f^{(n)}_{pnn}(\bm{r}_1) f^{(n)}_{pnn}(\bm{r}_2)$. 
Then we have
{\setlength\arraycolsep{0pt}
\begin{eqnarray}
&& \mathcal {A}_d =  8\int d\bm{r} f^{(n)}_{pn}(\bm{r}) e^{-\frac{\bm{r}'^2}{\sigma_d^2}} ,   \label{eq:Ad-initial}  \\
&& \mathcal {A}_{t} = 8^2 \int  d\bm{r}_1d\bm{r}_2 f^{(n)}_{ppn}(\bm{r}_1)  f^{(n)}_{ppn}(\bm{r}_2)
e^{-\frac{(\bm{r}'_1)^2+(\bm{r}'_2)^2}{\sigma_{t}^2}}.   \label{eq:AHe-initial} 
\end{eqnarray} }%

As in Ref.~\cite{fr2017acta,RQWang2021PRC}, we adopt $f^{(n)}_{pn}(\bm{r}) = \frac{1}{(\pi C R_f^2)^{1.5}} e^{-\frac{\bm{r}^2}{C R_f^2}}$ and
$f^{(n)}_{pnn}(\bm{r}_1) = \frac{1}{(\pi C_1 R_f^2)^{1.5}} e^{-\frac{\bm{r}_1^2}{C_1 R_f^2}}$, $f^{(n)}_{pnn}(\bm{r}_2) = \frac{1}{(\pi C_2 R_f^2)^{1.5}} e^{-\frac{\bm{r}_2^2}{C_2 R_f^2}}$,
where $R_f$ is the effective radius of the source system at the light nuclei freeze-out and $C$, $C_1$ and $C_2$ are distribution width parameters. 
Considering relations between $\bm{r}$, $\bm{r}_1$ and $\bm{r}_2$ with $\bm{x}_1$, $\bm{x}_2$ and $\bm{x}_3$, $C_1$ should be equal to $C/2$ 
and $C_2$ should be equal to $2C/3$.
So there is only one distribution width parameter $C$ to be determined.
In this article we set it to be 4, the same as that in Ref.~\cite{fr2017acta,RQWang2021PRC}. 

Considering instantaneous coalescence in the rest frame of $pn$-pair or $pnn$-cluster, i.e., $\Delta t'=0$, we get
\begin{eqnarray}
\bm{r} = \bm{r}' +(\gamma-1)\frac{\bm{r}'\cdot \bm{\beta}}{\beta^2}\bm{\beta}.
\end{eqnarray}
Substituting the above equation into Eqs.~(\ref{eq:Ad-initial}) and (\ref{eq:AHe-initial}) and integrating from relative coordinate variables, we can obtain 
{\setlength\arraycolsep{0pt}
\begin{eqnarray}
 \mathcal {A}_{d} &=& \frac{8\sigma_d^3}{(C R_f^2+\sigma_d^2) \sqrt{C (R_f/\gamma)^2+\sigma_d^2}},   \label{eq:Ad-fin}  \\
 \mathcal {A}_{t} &=& \frac{8\sigma_{t}^3}{(\frac{C}{2} R_f^2+\sigma_{t}^2) \sqrt{\frac{C}{2} (R_f/\gamma)^2+\sigma_{t}^2}}  \nonumber  \\
&&  \times \frac{8\sigma_{t}^3}{(\frac{2C}{3} R_f^2+\sigma_{t}^2) \sqrt{\frac{2C}{3} (R_f/\gamma)^2+\sigma_{t}^2}} .   \label{eq:AHe-fin}  
\end{eqnarray} }%


Recalling that $\sigma_d=\sqrt{\frac{8}{3}} R_d$ and $\sigma_{t}=R_{t}$,
where the root-mean-square charge radius of the deuteron $R_d$=2.1421 fm and that of the $t$ $R_{t}$=1.7591 fm \cite{radiiNPA2019},
we see that the gaussian width values $2\hbar c/\sigma_d$, $\sqrt{2}\hbar c/\sigma_{t}$ and $\sqrt{6}\hbar c/\sigma_{t}$ in Eqs.~(\ref{eq:Md}) and (\ref{eq:Mt}) are quite small.
So we can mathematically approximate the gaussian form of the kernel function $e^{-(\Delta \bm{p}')^2/\epsilon^2}$ as $(\sqrt{\pi} \epsilon)^3 \delta(\Delta \bm{p}')$, 
where $\epsilon$ is a small quantity.
Then we immediately obtain
{\setlength\arraycolsep{0.2pt}
\begin{eqnarray}
 \mathcal {M}_{d}(\bm{p}) 
&=&  (\frac{2\hbar c}{\sigma_d}\sqrt{\pi})^3
    \int d\bm{p}_1d\bm{p}_2 f^{(n)}_{pn}(\bm{p}_1,\bm{p}_2) \delta(\bm{p}'_{1}-\bm{p}'_{2})  \nonumber  \\
 && \times \delta(\displaystyle{\sum^2_{i=1}} \bm{p}_i-\bm{p}_d)    \nonumber   \\
 &=& (\frac{2\hbar c}{\sigma_d}\sqrt{\pi})^3 
    \int d\bm{p}_1d\bm{p}_2 f^{(n)}_{pn}(\bm{p}_1,\bm{p}_2) \gamma \delta(\bm{p}_{1}-\bm{p}_{2})  \nonumber  \\
 && \times \delta(\displaystyle{\sum^2_{i=1}} \bm{p}_i-\bm{p}_d)    \nonumber   \\
 &=& (\frac{\hbar c}{\sigma_d}\sqrt{\pi})^3 \gamma  f^{(n)}_{pn}(\frac{\bm{p}}{2},\frac{\bm{p}}{2}),  \label{eq:Md-fin}  
\end{eqnarray} }%
where $\gamma$ comes from $\Delta \bm{p}'=\frac{1}{\gamma}\Delta \bm{p}$.
Similarly we get
{\setlength\arraycolsep{0.2pt}
\begin{eqnarray}
 \mathcal {M}_{t} (\bm{p})
=  (\frac{\pi\hbar^2 c^2}{\sqrt{3}\sigma_{t}^2})^3 \gamma^2  f^{(n)}_{pnn}(\frac{\bm{p}}{3},\frac{\bm{p}}{3},\frac{\bm{p}}{3})  .     \label{eq:Mt-fin}  
\end{eqnarray} }%
The robustness of the above $\delta$ function approximation has been checked in our recent work~\cite{RQWang2021PRC}.

Substituting Eqs.~(\ref{eq:Ad-fin}-\ref{eq:Mt-fin}) into Eqs.~(\ref{eq:fd-AM}) and (\ref{eq:ft-AM}),
and ignoring correlations between protons and neutrons,
we finally have the momentum distributions of light nuclei as
{\setlength\arraycolsep{0.2pt}
\begin{eqnarray}
&& f_{d}(\bm{p}) = \frac{ 8(\sqrt{\pi}\hbar c)^3 g_{d} \gamma}{(C R_f^2+\sigma_d^2) \sqrt{C (R_f/\gamma)^2+\sigma_d^2}} 
                       f_{p}(\frac{\bm{p}}{2}) f_{n}(\frac{\bm{p}}{2}) ,     \label{eq:fd-final}  \\
&& f_{t}(\bm{p}) = \frac{8^2 (\pi\hbar^2 c^2)^3 g_{t} \gamma^2 }{3\sqrt{3}(\frac{C}{2} R_f^2+\sigma_{t}^2) \sqrt{\frac{C}{2} (R_f/\gamma)^2+\sigma_{t}^2} } \nonumber  \\
&&    \times  \frac{1}{(\frac{2C}{3} R_f^2+\sigma_{t}^2) \sqrt{\frac{2C}{3} (R_f/\gamma)^2+\sigma_{t}^2}} 
  f_{p}(\frac{\bm{p}}{3}) f_{n}(\frac{\bm{p}}{3}) f_{n}(\frac{\bm{p}}{3}). ~~~~~~    \label{eq:ft-final}  
\end{eqnarray} }%
From eqs.~(\ref{eq:fd-final}) and (\ref{eq:ft-final}), we can get the Lorentz invariant momentum distributions of light nuclei. 
We denote the invariant distribution $\dfrac{d^{2}N}{2\pi p_{T}dp_{T}dy}$ with $f^{inv}$ and at the midrapidity $y=0$ we have
\begin{align}
&f_{d}^{inv}(p_{T})=\frac{ 32(\sqrt{\pi}\hbar c)^3 g_{d} }{m_{d}(C R_f^2+\sigma_d^2) \sqrt{C (R_f/\gamma)^2+\sigma_d^2}}   \nonumber \\
& ~~~~~~~~~~~~~~~~~\times f_{p}^{inv}(\frac{p_{T}}{2}) f_{n}^{inv}(\frac{p_{T}}{2}) ,\label{eq:pt-d}\\
&f_{t}^{inv}(p_{T})=\frac{192\sqrt{3} (\pi\hbar^2 c^2)^3 g_{t} }{m_{t}^{2}(\frac{C}{2} R_f^2+\sigma_{t}^2) \sqrt{\frac{C}{2} (R_f/\gamma)^2+\sigma_{t}^2}(\frac{2C}{3} R_f^2+\sigma_{t}^2) } \nonumber  \\
&~~~~ \times  \frac{1}{ \sqrt{\frac{2C}{3} (R_f/\gamma)^2+\sigma_{t}^2}} 
f_{p}^{inv}(\frac{p_{T}}{3}) f_{n}^{inv}(\frac{p_{T}}{3}) f_{n}^{inv}(\frac{p_{T}}{3}).\label{eq:pt-t}
\end{align}
Eqs.~(\ref{eq:pt-d}) and (\ref{eq:pt-t}) show relationships of light nuclei with primordial nucleons in momentum space in the laboratory frame. They can be directly used to calculate the yields and $p_T$ spectra of light nuclei measured extensively as long as the nucleon Lorentz invariant momentum distributions are given.

\section{Results and discussions}    \label{Results}

In this section, we apply the deduced results in Sec.~\ref{model} to the midrapidity region of Au-Au collisions at the RHIC energies
to study production characteristics of  light (anti-)nuclei from the nucleon coalescence. 
First we give the $p_T$ distributions of final-state (anti-) protons and those at the kinetic freeze-out calculated by the SDQCM ~\cite{yanting work}. 
Then we present the results of $p_T$ distributions of light nuclei and antinuclei. 
Finally we show yields and several interesting yield ratios of different light (anti-)nuclei $d/p$, $\bar d/\bar p$, $t/p$, $\bar t/\bar p$, $d/p^2$, $\bar d/\bar p^2$, $tp/d^2$, etc., and discuss their properties as functions of the collision energy and the collision centrality.

\subsection{$p_T$ spectra of protons and antiprotons} 

The (anti-)nucleon $p_{T}$ distributions are necessary for computing $p_{T}$ distributions of light (anti-)nuclei in our method.
We use SDQCM to obtain invariant $p_{T}$ distributions of protons and antiprotons at final state as well as those at the kinetic freeze-out.
The detailed calculations for hadron production at the RHIC beam energy scan with the SDQCM can be found in our previous works ~\cite{Song2020PRC,Song2021PRC,yanting work}.

\begin{figure*}[htbp]
	\centering
	\includegraphics[width=0.8\linewidth]{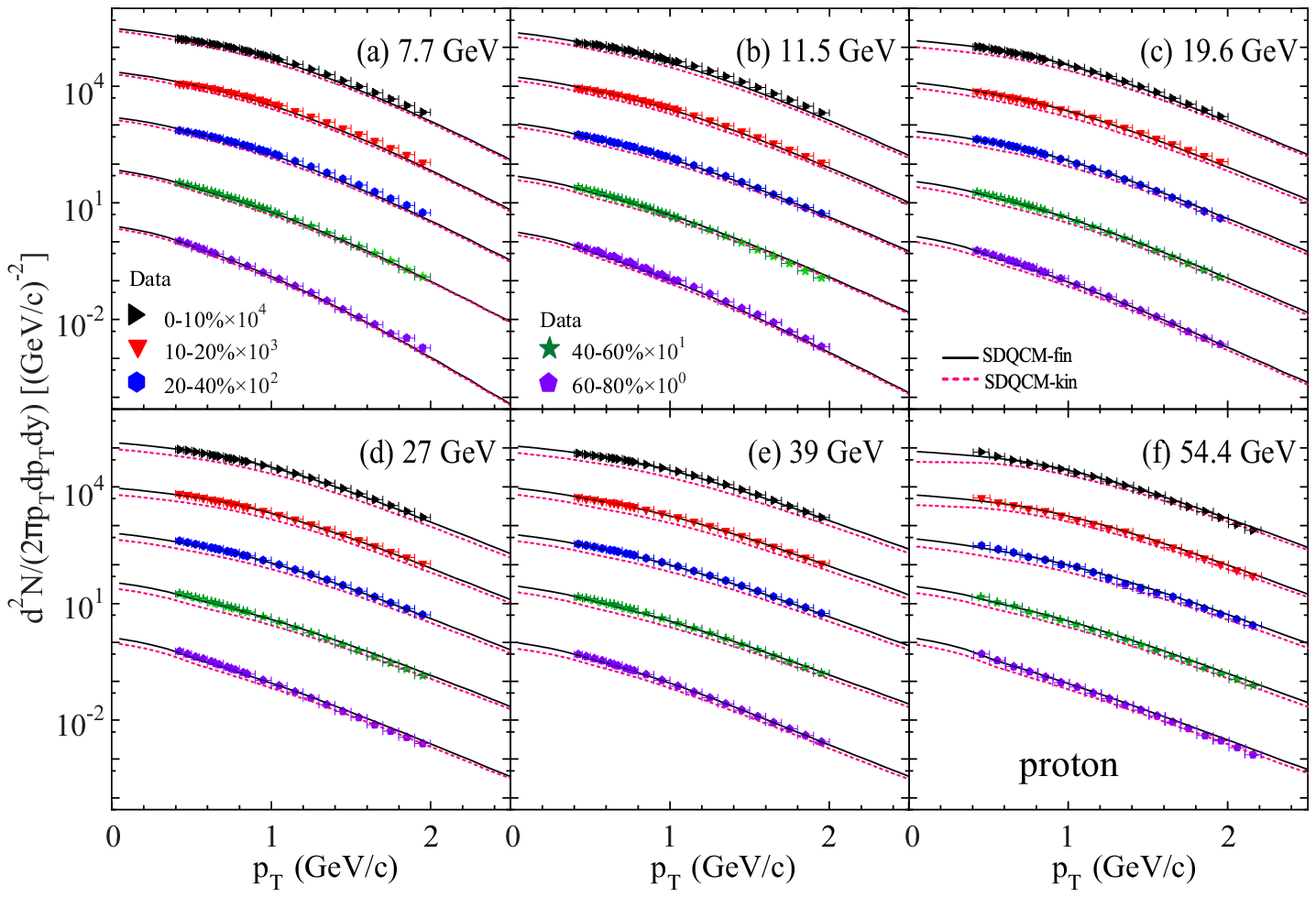}\\
    \centering
	\caption{Invariant $p_T$ spectra of protons at midrapidity in Au-Au collisions at $\sqrt{s_{NN}}=7.7,~11.5,~19.6,~27,~39,~54.4$ GeV in centralities $0-10\%$, $10-20\%$, $20-40\%$, $40-60\%$, and $60-80\%$. Filled symbols are experimental data from the STAR Collaboration~\cite{DWZhang2021NPASTAR,STAR:2017sal}. 
Solid lines are the results of fianl state protons and dashed lines are those at the kinetic freeze-out calculated by the SDQCM.}
	\label{fig:proton}
\end{figure*}

\begin{figure*}[htbp]
	\centering
	\includegraphics[width=0.8\linewidth]{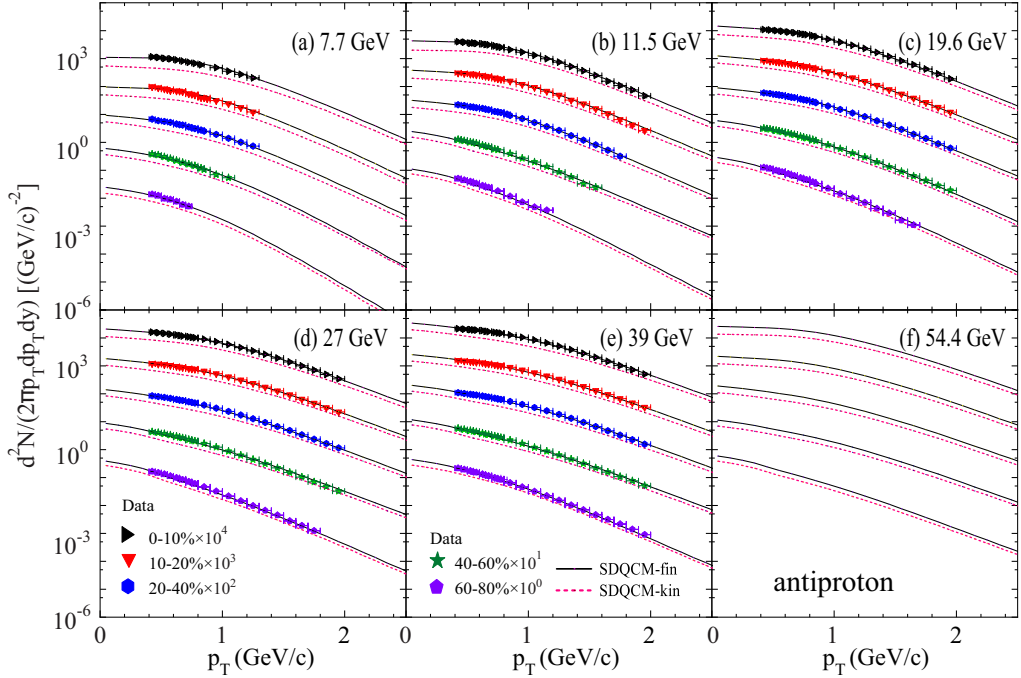}\\
	\centering
	\caption{Invariant $p_T$ spectra of antiprotons at midrapidity in Au-Au collisions at $\sqrt{s_{NN}}=7.7,~11.5,~19.6,~27,~39,~54.4$ GeV in centralities $0-10\%$, $10-20\%$, $20-40\%$, $40-60\%$, and $60-80\%$. Filled symbols are experimental data from the STAR Collaboration~\cite{DWZhang2021NPASTAR,STAR:2017sal}. Solid lines are the results of final state antiprotons and dashed lines are those at the kinetic freeze-out calculated by the SDQCM.}
	\label{fig:pbar}
\end{figure*}

Fig. \ref{fig:proton} and Fig. \ref{fig:pbar} show the invariant $p_T$ spectra of protons and antiprotons at midrapidity in Au-Au collisions at $\sqrt{s_{NN}}=7.7,~11.5,~19.6,~27,~39,~54.4$ GeV in centralities $0-10\%$, $10-20\%$, $20-40\%$, $40-60\%$, and $60-80\%$. 
Filled symbols are experimental data from the STAR Collaboration in Refs.~\cite{DWZhang2021NPASTAR,STAR:2017sal}. 
Solid lines are the results of final state protons calculated by the SDQCM, which describe the data well.
Dashed lines are the results of (anti-)protons at the kinetic freeze-out, which are just those we need for computing the production of light (anti-)nuclei.
The surplus of solid lines compared to dashed lines comes from the weak decays of hyperons after the kinetic freeze-out.

To see weak decay contaminations more clearly, we show the yield density $dN/dy$ of protons and that of antiprotons for final state ones and those corrected weak decays in Table \ref{tab:dNdy-p}. 
The contamination ratio from weak decays (WDC) is evaluated and the results are in fifth and eighth columns, from which one can see that the weak decay corrections exhibit explicit energy and centrality dependencies.
The contamination from weak decays for both protons and antiprotons becomes larger in more central collisions at the same collision energy because strangeness production is enhanced from peripheral to central collisions.
At the same centrality bin, WDC for protons becomes larger while for antiprotons it becomes smaller as the function of the colliding energy.
This is due to more newborn baryons and slightly decreasing strangeness production with the increasing energy.

\begin{table*}[htbp]
	\centering
	\caption{Yield densities $dN/dy$ of protons and antiprotons at midrapidity in Au-Au collisions at $\sqrt{s_{NN}}=7.7,~11.5,~19.6,~27,~39,~54.4$ GeV. 
Data in the third and sixth columns are from Ref.~\cite{STAR:2017sal}. QCM-fin at $\sqrt{s_{NN}}=54.4$ GeV denotes final state protons and antiprotons calculated by the SDQCM.
 QCM-cor in the forth and seventh columns denotes (anti-) protons corrected from weak decays calculated by the SDQCM. Weak decay contribution (WDC) ratio is evaluated and the results are in fifth and eighth columns.}
	\begin{tabular}{ccccccccc}
		\toprule
		\multirow{2}{*}{$\sqrt{s_{NN}}$}&\multirow{2}{*}{Centrality}&&$p$&&&&$\bar{p}$& \\ \cline{3-5}\cline{7-9}
		&&data&QCM-cor &WDC &&data&QCM-cor &WDC \\
		\hline
		\multirow{5}{*}{$7.7$ GeV}&$0-10\%$&$50.2\pm5.6$&$41.0$&$18.2\%$&&$0.36\pm0.05$&$0.17$&$52.5\%$\\
		&$10-20\%$&$33.4\pm3.7$&$27.5$&$17.5\%$&&$0.26\pm0.03$&$0.13$&$50.7\%$ \\
		&$20-40\%$&$19.5\pm2.2$&$16.3$&$16.5\%$&&$0.17\pm0.02$&$0.09$& $46.2\%$\\ 	
		&$40-60\%$&$7.4\pm0.8$&$6.3$&$14.9\%$&&$0.08\pm0.01$&$0.05$&$40.0\%$ \\		
		&$60-80\%$&$2.1\pm0.3$&$1.8$&$12.9\%$&&$0.026\pm0.003$&$0.016$&$36.7\%$\\
		&&&&&&&&\\
			
		\multirow{5}{*}{$11.5$ GeV}&$0-10\%$&$39.6\pm4.8$&$30.2$&$23.7\%$&&$1.4\pm0.2$&$0.7$&$48.4\%$\\
		&$10-20\%$&$26.1\pm3.1$&$20.1$&$23.0\%$&&$0.9\pm0.1$&$0.5$&$46.7\%$\\
		&$20-40\%$&$14.8\pm1.8$&$11.5$&$22.3\%$&&$0.6\pm0.1$&$0.3$&$44.4\%$\\
		&$40-60\%$&$5.8\pm0.7$&$4.5$&$20.9\%$&&$0.27\pm0.04$&$0.17$&$38.6\%$\\
		&$60-80\%$&$1.6\pm0.2$&$1.3$&$17.7\%$&&$0.10\pm0.02$&$0.06$&$33.5\%$ \\
		&&&&&&&&\\
		
		\multirow{5}{*}{$19.6$ GeV}&$0-10\%$&$31.8\pm4.2$&$22.1$&$30.3\%$&&$3.8\pm0.5$&$2.1$&$45.3\%$\\
		&$10-20\%$&$21.9\pm2.9$&$15.5$&$29.5\%$&&$2.7\pm0.4$&$1.5$&$44.5\%$ \\
		&$20-40\%$&$11.9\pm1.6$&$8.4$&$28.9\%$&&$1.7\pm0.3$&$1.0$& $42.2\%$\\
		&$40-60\%$&$4.6\pm0.6$&$3.3$&$26.9\%$&&$0.8\pm0.1$&$0.5$&$37.4\%$\\
		&$60-80\%$&$1.3\pm0.2$&$1.0$&$25.0\%$&&$0.27\pm0.04$&$0.18$&$33.2\%$ \\
		&&&&&&&&\\
		
		\multirow{5}{*}{$27$ GeV}&$0-10\%$&$29.1\pm3.5$&$19.4$&$33.3\%$&&$5.6\pm0.7$&$3.2$&$43.2\%$\\
		&$10-20\%$&$19.4\pm2.3$&$13.3$&$31.2\%$&&$4.0\pm0.5$&$2.3$&$41.7\%$   \\
		&$20-40\%$&$10.9\pm1.3$&$7.6$&$30.8\%$&&$2.5\pm0.3$&$1.5$& $40.1\%$ \\
		&$40-60\%$&$4.4\pm0.6$&$3.1$&$29.6\%$&&$1.1\pm0.2$&$0.7$&$36.7\%$    \\
		&$60-80\%$&$1.3\pm0.2$&$0.2$&$25.4\%$&&$0.36\pm0.04$&$0.25$& $31.4\%$ \\
		&&&&&&&&\\
		
		\multirow{5}{*}{$39$ GeV}&$0-10\%$&$24.6\pm2.7$& $16.2$& $34.1\%$&&$8.0\pm1.0$&$4.6$& $42.3\%$  \\
		&$10-20\%$&$17.3\pm1.9$&$11.5$&$33.5\%$&&$5.4\pm0.7$& $3.2$&$41.1\%$   \\
		&$20-40\%$&$9.9\pm1.1$&$6.7$&$32.2\%$&&$3.4\pm0.4$&$2.1$& $39.3\%$  \\
		&$40-60\%$&$3.9\pm0.4$&$2.7$&$30.2\%$&&$1.5\pm0.2$&$1.0$&$35.2\%$   \\
		&$60-80\%$&$1.1\pm0.2$&$0.8$&$27.5\%$&&$0.49\pm0.06$&$0.33$& $32.2\%$ \\ 
		&&&&&&&&\\
		&&QCM-fin&&&&QCM-fin&&\\
		\multirow{5}{*}{$54.4$ GeV}&$0-10\%$&$23.6$&$15.3$&$35.2\%$&&$9.9$& $5.8$& $41.3\%$ \\
		&$10-20\%$&16.2&$10.5$&$34.8\%$&&$6.8$&$4.0$&$41.0\%$  \\
		&$20-40\%$&8.7& $5.7$&$34.7\%$&&$4.1$&$2.5$& $39.8\%$  \\
		&$40-60\%$&3.5&$2.4$&$30.8\%$&&$1.8$&$1.2$&$35.3\%$ \\
		&$60-80\%$&1.1& $0.7$&$29.7\%$&&$0.6$&$0.4$& $32.8\%$ \\ 
		\hline\hline
		\label{tab:dNdy-p}
	\end{tabular}
\end{table*}
\subsection{$p_T$ spectra of light nuclei and antinuclei} 

According to Eqs. (\ref{eq:pt-d}) and (\ref{eq:pt-t}), the $p_{T}$ distributions of deuterons and tritons can be computed with the proton $p_{T}$ distributions shown in Fig. \ref{fig:proton}. 
The isospin symmetry is adopted, i.e., we assume the $p_{T}$ distribution of the neutron is the same with that of the proton.
The effective radius of the hadronic system $R_{f}$ is characterized by the rapidity density of charged particles $dN_{ch}/dy$ as $R_{f}=a*(dN_{ch}/dy)^{1/3}$~\cite{Rf2005,Rf2016ALICE}, and $a$ is a free parameter. 
In the current article, $a=0.58$ for both $d$ and $\bar d$, and $a=0.55$ for both $t$ and $\bar t$. 
The slightly lower value of $a$ for $t,~\bar t$ may indicate their earlier freezeout compared to $d,~\bar d$ in our model.
These values are comparable to those we previously adopted in Ref.~\cite{RQWang2021PRC}.
With the data of  $dN_{ch}/dy$ in Ref.~\cite{STAR:2017sal}, we get the value of $R_f$ and then we can compute $p_{T}$ distributions of $d$, $\bar d$, $t$ and $\bar t$.

\begin{figure*}
\centering
\includegraphics[width=0.88\linewidth]{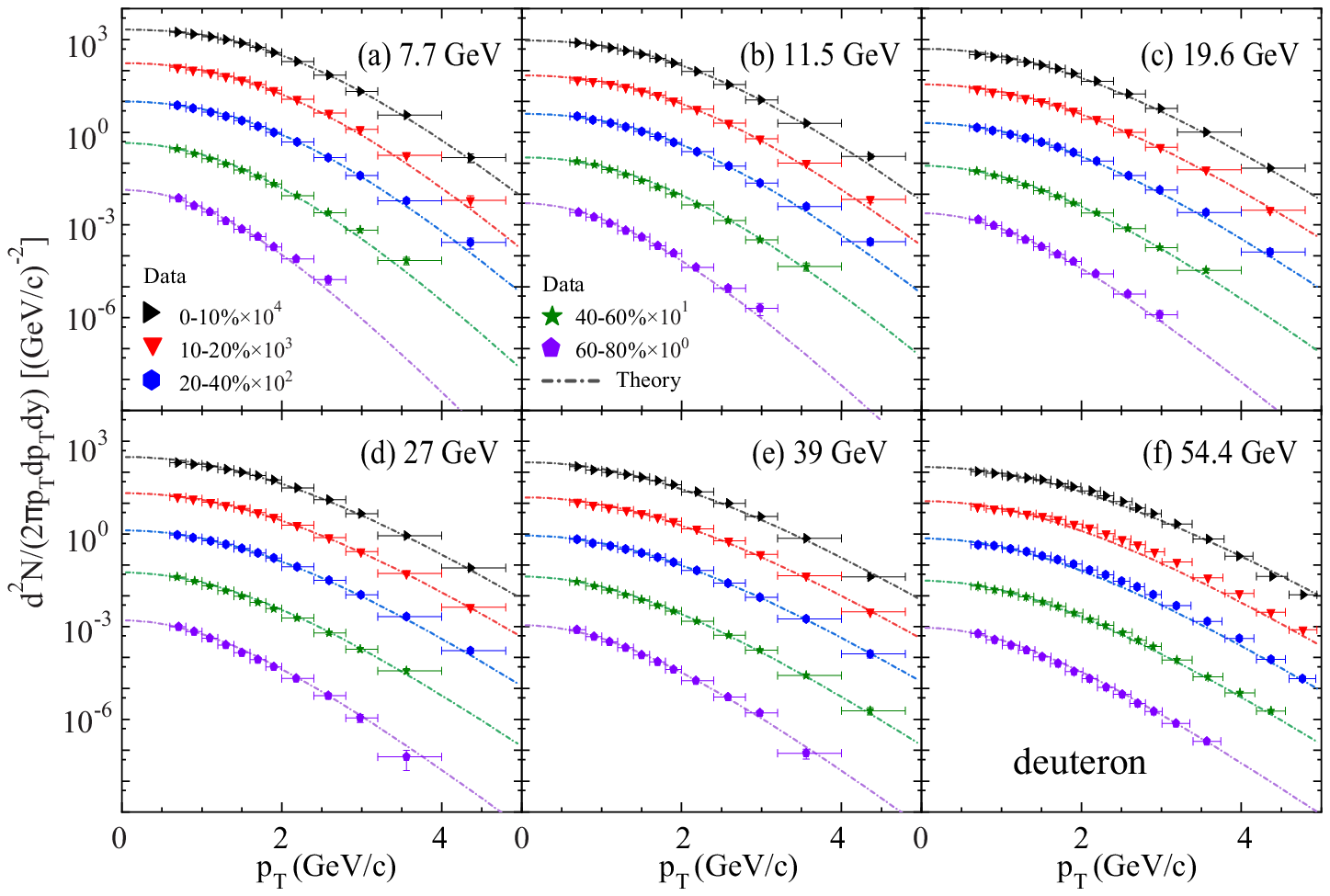}\\
\centering
\caption{Invariant $p_T$ spectra of deuterons at midrapidity in Au-Au collisions in different centralities at $\sqrt{s_{NN}}=7.7,~11.5,~19.6,~27,~39,~54.4$ GeV.
Filled symbols are the data~\cite{B2d2019PRCSTAR,DWZhang2021NPASTAR}.
Dashed-dotted lines are our theoretical results. Spectra for different centralities are scaled by different factors for clarity.}
 \label{fig:deuteron}
\end{figure*}

\begin{figure*}
\centering
\includegraphics[width=0.88\linewidth]{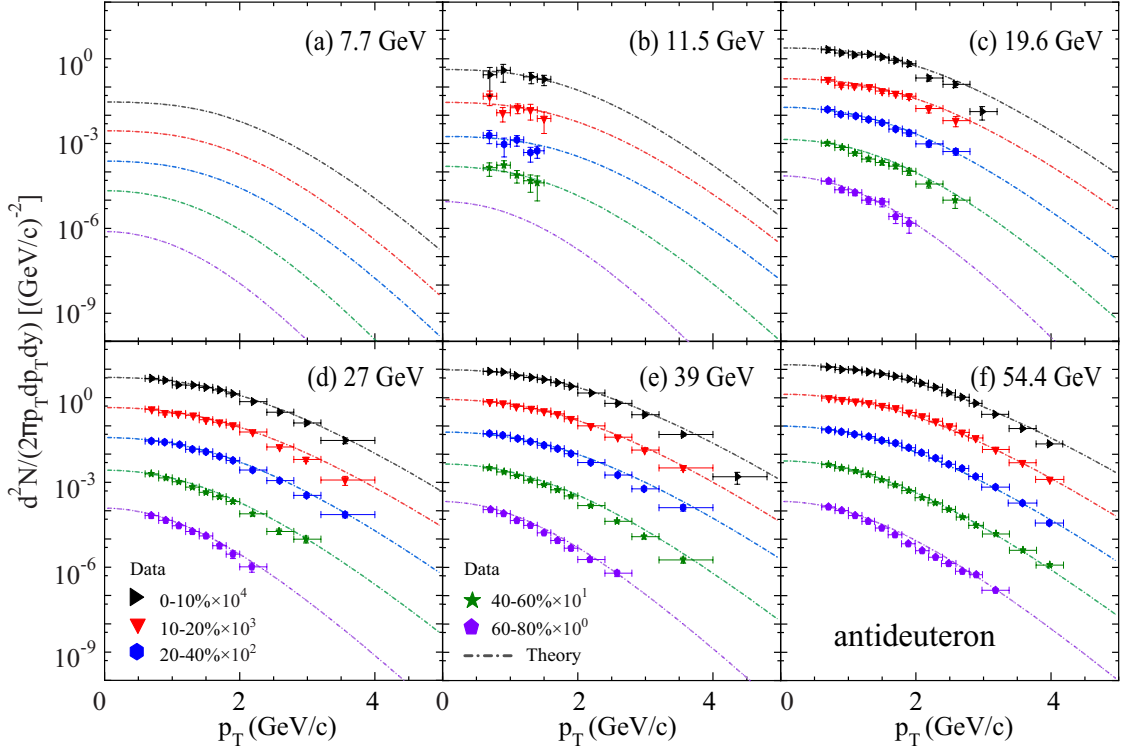}\\
\centering
\caption{Invariant $p_T$ spectra of antideuterons at midrapidity in Au-Au collisions in different centralities at $\sqrt{s_{NN}}=7.7,~11.5,~19.6,~27,~39,~54.4$ GeV. Filled symbols are the data~\cite{B2d2019PRCSTAR,DWZhang2021NPASTAR}. Dashed-dotted lines are our theoretical results. The spectra for different centralities are scaled  by different factors for clarity.}
\label{fig:dbar}
\end{figure*}

Figs.~\ref{fig:deuteron} and \ref{fig:dbar} show $p_T$ spectra for $d$ and $\bar d$ at midrapidity in Au-Au collisions in $0-10\%$, $10-20\%$, $20-40\%$, $40-60\%$, and $60-80\%$ centralities at $\sqrt{s_{NN}}=7.7,~11.5,~19.6,~27,~39,~54.4$ GeV.
Filled symbols are the data from STAR Collaboration~\cite{B2d2019PRCSTAR,DWZhang2021NPASTAR}.
Dashed-dotted lines are our theoretical results. The spectra for different centralities are scaled by different factors for clarity as shown in the figures.
From Figs.~\ref{fig:deuteron} and \ref{fig:dbar}, one can see the coalescence model can well reproduce the available data for both $d$ and $\bar d$ from central to peripheral Au-Au collisions at the beam energy scan energies.

\begin{figure*}
\centering
\includegraphics[width=0.88\linewidth]{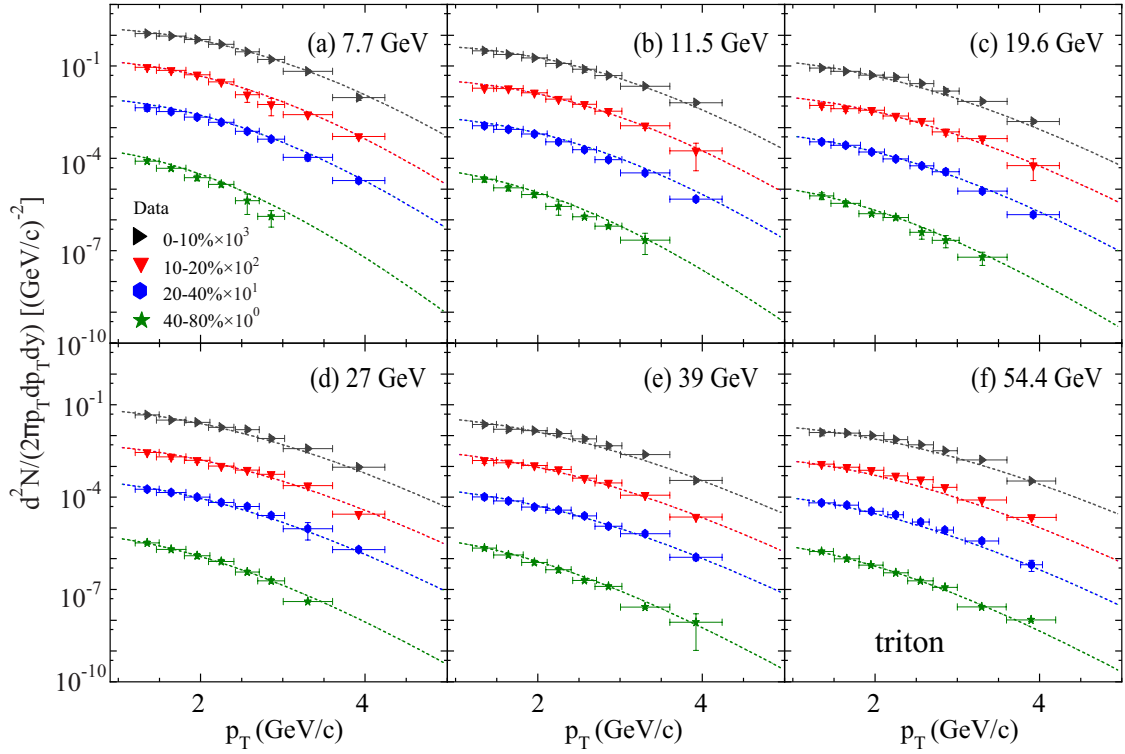}\\
\centering
\caption{Invariant $p_T$ spectra of tritons at midrapidity in Au-Au collisions in different centralities at $\sqrt{s_{NN}}=7.7,~11.5,~19.6,~27,~39,~54.4$ GeV. Filled symbols are the data~\cite{DWZhang2021NPASTAR}. Different dashed lines are theoretical results. Spectra in different centralities are scaled by different factors for clarity.}
\label{fig:triton}
\end{figure*}

\begin{figure*}
\centering
\includegraphics[width=0.87\linewidth]{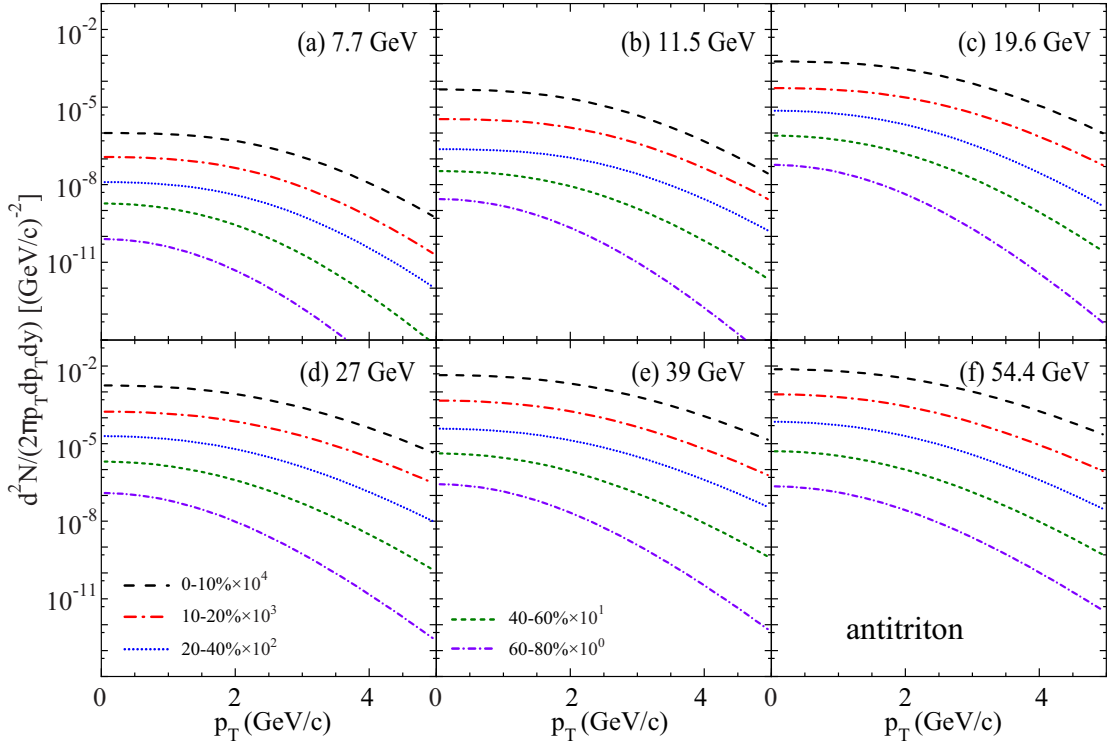}\\
\centering
\caption{Predictions for invariant $p_T$ spectra of antitritons at midrapidity in Au-Au collisions in different centralities at $\sqrt{s_{NN}}=7.7,~11.5,~19.6,~27,~39,~54.4$ GeV. Spectra for different centralities denoted by different lines are scaled by different factors for clarity.}
\label{fig:tbar}
\end{figure*}

Fig.~\ref{fig:triton} shows $p_T$ spectra of tritons at midrapidity in Au-Au collisions in $0-10\%$, $10-20\%$, $20-40\%$ and $40-80\%$ centralities at $\sqrt{s_{NN}}=7.7,~11.5,~19.6,~27,~39,~54.4$ GeV.
Filled symbols are the data from STAR Collaboration~\cite{DWZhang2021NPASTAR}.
Dashed lines are our theoretical results, which agree well the the available data. The spectra for different centralities are scaled by different factors for clarity as shown in the figure.
We also predict the invariant $p_T$ spectra of antitritons and the results are in Fig.~\ref{fig:tbar}. 

The consistency between the theoretical results from the coalescence model and the data for $d$, $\bar d$, $t$ and $\bar t$ in Figs.~\ref{fig:deuteron}-\ref{fig:tbar} show the dominant role of the coalescence mechanism in describing the production of light nuclei and antinuclei at these RHIC energies.

\subsection{Yield densities $dN/dy$ of light nuclei and antinuclei} 

After integrating over the $p_T$, we can get the rapidity yield densities of light (anti-)nuclei.
Table \ref{tab:dNdy} shows our results of  $d$, $\bar d$, $t$, and $\bar t$ in Au-Au collisions at midrapidity in different centralities at $\sqrt{s_{NN}}=7.7,~11.5,~19.6,~27,~39,~54.4$ GeV. 
Data with errors are from Refs.~\cite{B2d2019PRCSTAR,DWZhang2021NPASTAR}.
Our results are consistent with the available data.
$dN/dy$ becomes larger for both light nuclei and antinuclei from peripheral to central collisions at the same collision energy.
This is due to the more energy deposited in the reaction region in more central collisions.
For the same centrality, $dN/dy$ of light nuclei decreases gradually, while for antinuclei it increases with the increasing collision energy.
This is related with the net nucleons from the colliding heavy nuclei.
It is easier for them to stop in the midrapidity region to form light nucei in lower collision energies.

\begin{table*}
	\centering
	\caption{Yield densities $dN/dy$ of $d$, $\bar d$, $t$, and $\bar t$ at midrapidity in Au-Au collisions in different centralities at $\sqrt{s_{NN}}=7.7,~11.5,~19.6,~27,~39,~54.4$ GeV. Data are from Refs.~\cite{B2d2019PRCSTAR,DWZhang2021NPASTAR}.}
	\begin{tabular}{cccccccccccc}
		\toprule
		\multirow{2}{*}{$\sqrt{s_{NN}}$}&\multirow{2}{*}{Centrality}&\multicolumn{2}{c}{$d$}&&\multicolumn{2}{c}{$\bar{d}$}&&\multicolumn{2}{c}{$t$}&&$\bar{t}$ \\ \cline{3-4}\cline{6-7}\cline{9-10}\cline{12-12}
		&&data &theory$~~~~$ &&data &theory$~~~~$ &&data &theory$~~~~$&&theory \\
		\hline
		&&$\times10^{-2}$&$\times10^{-2}$&&$\times10^{-5}$&$\times10^{-5}$&&$\times10^{-3}$&$\times10^{-3}$&&$\times 10^{-9}$ \\
		\multirow{5}{*}{$7.7$ GeV}&$0-10\%$&$140.99\pm0.41\pm10.97$&$142.52$&&$-$&$2.43$&&$21.64$&$22.06$&&$1.48$\\
		&$10-20\%$&$93.87\pm0.32\pm7.92$&$96.73$&&$-$&$1.94$&&$15.76$&$16.51$&&$1.41 $ \\	
		&$20-40\%$&$49.06\pm0.16\pm5.38$&$51.86$&&$-$&$1.46$&&$7.30$&$9.23$&&$1.33$ \\ 
		&$40-60\%$&$15.48\pm0.09\pm2.92$&$16.52$&&$-$&0.89&&\multirow{2}{*}{$1.25$}&\multirow{2}{*}{$1.44$}&&$1.25$ \\
		&$60-80\%$&$3.13\pm0.05\pm0.91$& $3.28$&&$-$&$0.22$&&&&&$0.36$\\
		\hline
		
		&&$\times10^{-2}$&$\times10^{-2}$&&$\times10^{-4}$&$\times10^{-4}$&&$\times10^{-3}$&$\times10^{-3}$&&$\times 10^{-8}$ \\
		\multirow{5}{*}{$11.5$ GeV}&$0-10\%$&$63.05\pm0.14\pm4.55$&$65.11$&&$3.29\pm0.63\pm1.10$&$3.21$&&$5.83$&$6.07$&&$6.50 $ \\
		&$10-20\%$&$41.02\pm0.11\pm3.39$&$42.46$&&$1.92\pm0.32\pm0.57$&$2.31$&&$4.00$& $4.23$&&$4.94$  \\
		&$20-40\%$&$21.92\pm0.06\pm2.23$&$22.32$&&$1.05\pm0.17\pm0.34$&$1.42$&&$1.96$& $2.34$ &&$3.28$  \\ 
		&$40-60\%$&$6.73\pm0.03\pm1.17$&$6.96$&&$-$&$0.84$&&\multirow{2}{*}{$0.34$}&\multirow{2}{*}{$0.34$}&&$3.01 $ \\
		&$60-80\%$&$1.31\pm0.02\pm0.40$& $1.38$&&$-$&$0.29$&&&&&$1.48$\\
		\hline
		
		&&$\times10^{-2}$&$\times10^{-2}$&&$\times10^{-4}$&$\times10^{-4}$&&$\times10^{-4}$&$\times10^{-4}$&&$\times 10^{-7}$ \\
		\multirow{5}{*}{$~~19.6$ GeV$~~$}&$0-10\%$&$27.45\pm0.06\pm2.04$&$29.12$&&$17.88\pm0.52\pm3.14$& $20.45$&&$15.70$&$16.43$&&$8.33$ \\
		&$10-20\%$&$18.78\pm0.05\pm1.57$&$20.09$&&$13.16\pm0.45\pm2.36$&$15.38$&&$10.20$&$11.98$&&$6.92$     \\
		&$20-40\%$&$9.73\pm0.03\pm1.00$&$10.05$&&$10.33\pm0.27\pm1.87$&$11.44$&&$5.37$&$6.23$&&$6.73$     \\ 
		&$40-60\%$&$3.20\pm0.01\pm0.55$&$3.30$&&$5.48\pm0.20\pm1.15$&$6.65$&&\multirow{2}{*}{$0.90$}&\multirow{2}{*}{$0.95$}&&$6.10$  \\
		&$60-80\%$&$0.68\pm0.007\pm0.21$&$0.67$&&$2.07\pm0.14\pm0.70$&$2.23$&&&&&$2.80$\\
		\hline
		
		&&$\times10^{-2}$&$\times10^{-2}$&&$\times10^{-4}$&$\times10^{-4}$&&$\times10^{-4}$&$\times10^{-4}$&&$\times 10^{-6}$ \\
		\multirow{5}{*}{$27$ GeV}&$0-10\%$&$18.44\pm0.04\pm1.28$&$19.57$&&$41.35\pm0.54\pm4.63$& $44.34$&&$7.98$& $8.35$&&$2.59$  \\
		&$10-20\%$&$12.83\pm0.03\pm1.05$&$12.98$&&$32.35\pm0.47\pm3.85$& $35.17$&&$5.07$& $5.71$&&$2.35$      \\
		&$20-40\%$&$6.84\pm0.01\pm0.70$&$7.05$&&$23.03\pm0.28\pm2.79$& $24.68$& &$3.17$& $3.38$&&$2.09$         \\ 
		&$40-60\%$&$2.33\pm0.009\pm0.43$&$2.45$&&$11.48\pm0.21\pm2.45$&$12.92$&&\multirow{2}{*}{$0.49$}&\multirow{2}{*}{$0.59$}&&$1.55$\\
		&$60-80\%$&$0.49\pm0.004\pm0.17$&$0.52$&&$3.33\pm0.11\pm1.23$&$3.87$&&&&&$0.60$ \\
		\hline
		
		&&$\times10^{-2}$&$\times10^{-2}$&&$\times10^{-4}$&$\times10^{-4}$&&$\times10^{-4}$&$\times10^{-4}$&&$\times 10^{-6}$ \\
		\multirow{5}{*}{$39$ GeV}&$0-10\%$&$12.73\pm0.02\pm0.95$&$13.27$&&$79.96\pm0.46\pm6.35$& $85.57$&&$4.21$& $4.59$&&$6.60$ \\
		&$10-20\%$&$8.78\pm0.01\pm0.69$&$9.20$&&$62.39\pm0.40\pm4.60$&$64.75$&&$3.01$& $3.32$&&$5.80$        \\
		&$20-40\%$&$4.81\pm0.008\pm0.48$&$5.03$&&$41.24\pm0.23\pm4.11$& $42.81$&&$1.68$& $1.92$&&$4.46$          \\ 
		&$40-60\%$&$1.72\pm0.004\pm0.30$&$1.82$&&$19.24\pm0.15\pm3.26$& $22.45$  &&\multirow{2}{*}{$0.36$}&\multirow{2}{*}{$0.37$}&&$3.33$\\
		&$60-80\%$&$0.37\pm0.002\pm0.12$&$0.37$&&$5.50\pm0.09\pm1.80$& $6.77$&&&&&$1.33$ \\
		\hline
		
		&&$\times10^{-2}$&$\times10^{-2}$&&$\times10^{-2}$&$\times10^{-2}$&&$\times10^{-4}$&$\times10^{-4}$&&$\times 10^{-5}$ \\
		\multirow{5}{*}{$54.4$ GeV}&$0-10\%$&$10.28$&$9.58$&&$1.21$&$1.26$&&$2.67$&$2.47$&&$1.22$   \\
		&$10-20\%$&$7.07$&$7.15$&&$0.93$&$0.93$&&$2.36$&$2.14$&&$0.94$   \\
		&$20-40\%$&$3.89$&$3.67$&&$0.57$&$0.59$&&$1.29$&$1.16$&&$0.69$  \\ 
		&$40-60\%$&$1.40$&$1.27$&&$0.28$&$0.28$&&\multirow{2}{*}{$0.25$}&\multirow{2}{*}{$0.21$}&&$0.39$\\
		&$60-80\%$&$0.31$ & $0.28$&&$0.07$&$0.08$&&&&&$0.13$   \\
		\hline\hline
		\label{tab:dNdy}
	\end{tabular}
\end{table*}

\subsection{Yield ratios of light nuclei and antinuclei}

Yield ratios of light (anti-)nuclei are more sensitive probes for the production mechanism and exhibit some interesting behaviors as functions of the collision energy and the collision centrality.
In this subsection, we systematically study different kinds of yield ratios.

\begin{figure}[htbp]
	\centering
	\includegraphics[width=1.0\linewidth]{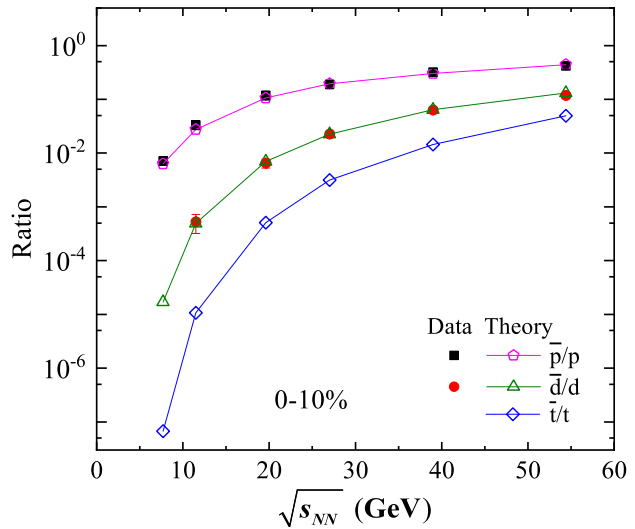}
	\caption{Energy dependence of $\bar{p}/p$, $\bar{d}/d$ and $\bar{t}/t$ in the most central 0-10\% Au-Au collisions.
    Data of $\bar{p}/p$ and $\bar{d}/d$ denoted by filled symbols are from Ref.~\cite{B2d2019PRCSTAR}. Open triangles and diamonds connected with lines to guide the eye are the theoretical results for $\bar{d}/d$ and $\bar{t}/t$, respectively.}
	\label{fig:dbard-010}
\end{figure}

Fig. \ref{fig:dbard-010} shows the ratios of antiparticles to particles $\bar{p}/p$, $\bar{d}/d$ and $\bar{t}/t$ in the most central $0-10\%$ centrality in Au-Au collisions at $\sqrt{s_{NN}}=7.7,~11.5, ~19.6,~27,~39,~54.4$ GeV. 
Filled squares are the data of $\bar{p}/p$ from the STAR Collaboration~\cite{B2d2019PRCSTAR}, and filled circles with error bars are the data of $\bar{d}/d$~\cite{B2d2019PRCSTAR}.
Open pentagons and triangles connected with lines to guide the eye are the theoretical results for $\bar{p}/p$ and $\bar{d}/d$, which agree well with the data.
Open diamonds connected with lines to guide the eye are the theoretical predictions for $\bar{t}/t$.
All these antiparticle-to-particle ratios increase and exhibit a distinct hierarchy with different constituent (anti-)nucleon numbers as the function of the collision energy. 
This is due to the decrease of net baryon density with the increasing energy. At very high collision energy such as at those at the LHC, it can be considered that the net baryon density is close to zero and all these ratios approach to one and their hierarchy with different constituent (anti-)nucleons disappear.

\begin{figure}[htbp]
	\centering
	\includegraphics[width=1.0\linewidth]{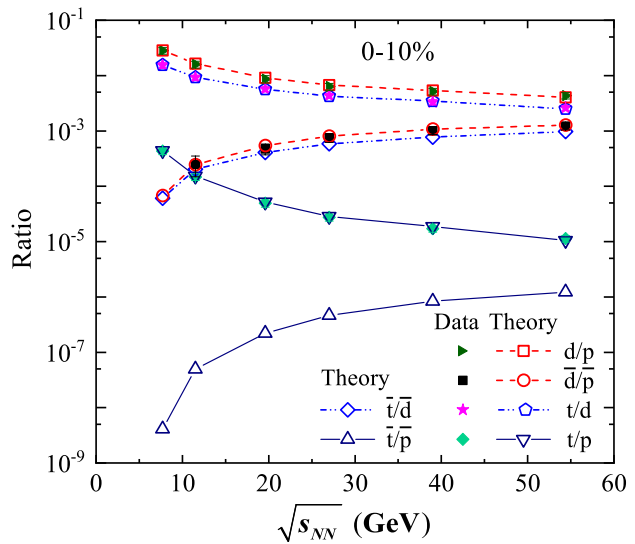}\\
	\centering
	\caption{Energy dependence of ratios $d/p$, $t/p$, $t/d$, $\bar{d}/\bar{p}$, $\bar{t}/\bar{p}$ and $\bar{t}/\bar{d}$ in the most central 0-10\% Au-Au collisions. Filled symbols are the data~\cite{STAR:2017sal,B2d2019PRCSTAR,DWZhang2021NPASTAR}. Open symbols connected with different lines to guide the eye are the theoretical results.}
	\label{fig:dptdtp-010}
\end{figure}

Fig. \ref{fig:dptdtp-010} shows the energy dependence of two-particle ratios $d/p$, $t/p$, $t/d$ and the corresponding antiparticle ratios $\bar d/\bar p$, $\bar t/\bar p$, $\bar t/\bar d$ in the most central $0-10\%$ centrality in Au-Au collisions at $\sqrt{s_{NN}}=7.7,~11.5,~19.6,~27,~39,~54.4$ GeV. 
Filled symbols are the data in Refs.~\cite{STAR:2017sal,B2d2019PRCSTAR,DWZhang2021NPASTAR}.
Open symbols connected with different lines to guide the eye are our results.
The two-particle ratios $d/p$, $t/p$, $t/d$ decrease while two-antiparticle ratios $\bar d/\bar p$, $\bar t/\bar p$, $\bar d/\bar p$ increase as the function of $\sqrt{s_{NN}}$. 
With the increasing $\sqrt{s_{NN}}$, the net nucleons stopped in the midrapidity region decreases while the energy deposited to create antinucleons increase.
This will enhance the antinucleon rapidity density and suppress the nucleon rapidity density.
These two-particle ratios are related with the nucleon density, and two-antiparticle ratios are related with the antinucleon density.
So they have different behaviors as the function of $\sqrt{s_{NN}}$.
Values of $d/p$ ($\bar d/\bar p$) are comparable to $t/d$ ($\bar d/\bar p$), and they are much larger than those of $t/p$ ($\bar t/\bar p$).
This is due to that $d/p$ and $t/d$ are proportional to the nucleon density while $t/p$ is proportional to the square of the nucleon density.
The similar case holds for two-antiparticle ratios.

\begin{figure*}[htbp]
	\centering
			\includegraphics[width=0.7\linewidth]{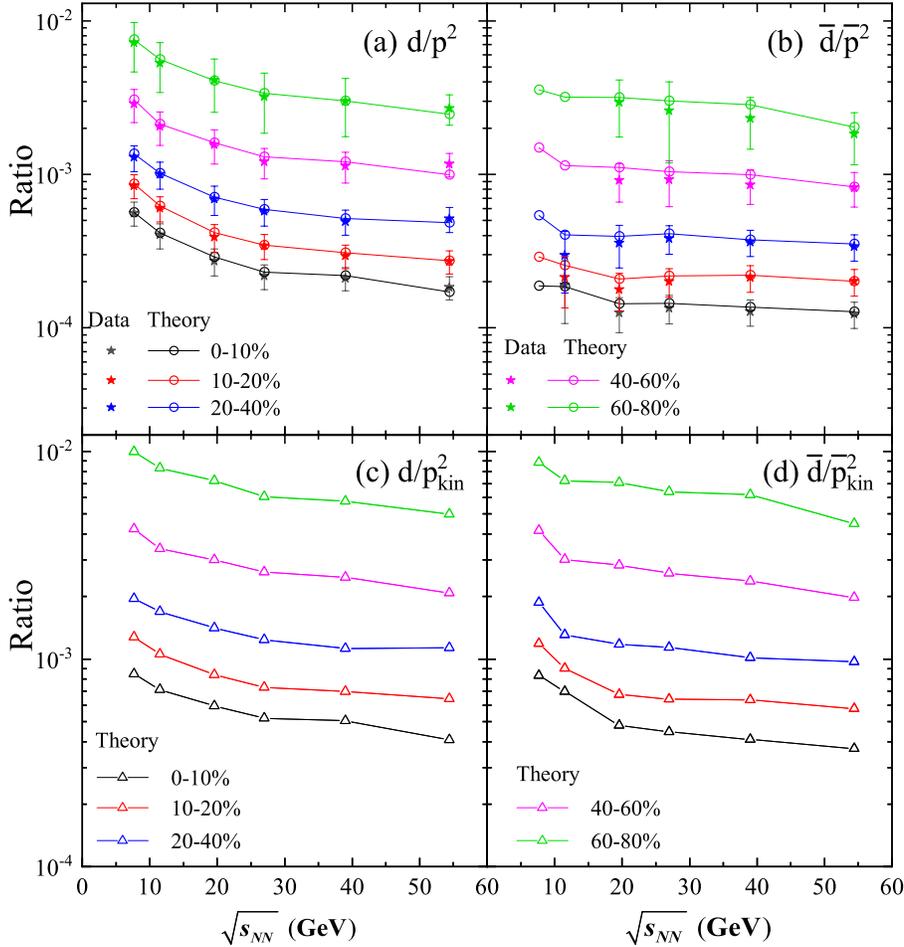}\\
	\centering
	\caption{Energy dependence of (a) $d/p^{2}$, (b) $\bar{d}/\bar{p}^{2}$, (c) $d/p^{2}_{kin}$ and (d) $\bar{d}/\bar{p}^{2}_{kin}$ at midrapidity in Au-Au collisions in different centralities. Data denoted by filled symbols with error bars are obtained according to the final-state proton and deuteron yields from STAR Collaboration \cite{B2d2019PRCSTAR,DWZhang2021NPASTAR,STAR:2017sal}. 
Open circles and triangles connected with lines to guide the eye are the theoretical results with final-state protons and kinetic freeze-out ones, respectively.}
	\label{fig:dp2-snn}
\end{figure*}

Fig. \ref{fig:dp2-snn} (a) and (b) show the energy dependence of ratios $d/p^{2}$ and $\bar{d}/\bar{p}^{2}$, respectively, in Au-Au collisions in $0-10\%$, $10-20\%$, $20-40\%$, $40-60\%$, $60-80\%$ centralities. 
Both  $d/p^{2}$ and $\bar{d}/\bar{p}^{2}$ decrease with the increase of $\sqrt{s_{NN}}$, which is very different from the previous $d/p$ and $\bar{d}/\bar{p}$. 
Note that $d/p^{2}$ and $\bar{d}/\bar{p}^{2}$ represent the probability of any nucleon-pair coalescencing into a deuteron and that of  any antinucleon-pair coalescencing into an antideuteron.
They do not depend on the absolute (anti-)nucleon numbers or the (anti-)nucleon rapidity densities, but are sensitive to the fundamental production mechanism. 
It is more difficult for any (anti-)nucleon-pair to recombining into (anti-)deuteron in larger hadronic system.
So $d/p^{2}$ and $\bar{d}/\bar{p}^{2}$ decrease with increasing $\sqrt{s_{NN}}$.

We want to point out that (anti-)protons in the ratios mentioned above are referred to those final-state ones including those from hyperon weak decays.
As is well known, (anti-)nucleons taking part in forming light (anti-)nuclei are those created before the kinetic freeze-out, not including those from hyperon weak decays.
To probe the production properties more directly, one should use (anti-)nucleons excluding hyperon weak decay contaminations, i.e., those at the kinetic freeze-out, to construct ratios.
Here, we present $d/p^{2}_{kin}$ and $\bar{d}/\bar{p}^{2}_{kin}$ in fig. \ref{fig:dp2-snn} (c) and (d) where the subscript $kin$ denote (anti-)protons at the kinetic freeze-out.
It can be found that $d/p^{2}_{kin}$ and $\bar{d}/\bar{p}^{2}_{kin}$ almost coincide with each other after correcting the weak decays of (anti-)protons from (anti-)hyperons.
This further indicate that the intrinsic dynamics of two nucleons recombining into a deuteron is similar with that of two antinucleons recombining into an antideuteron.

\subsection{Multi-particle yield correlation $tp/d^2$}

\begin{figure*}[htbp]
	\centering
	\includegraphics[width=0.95\linewidth]{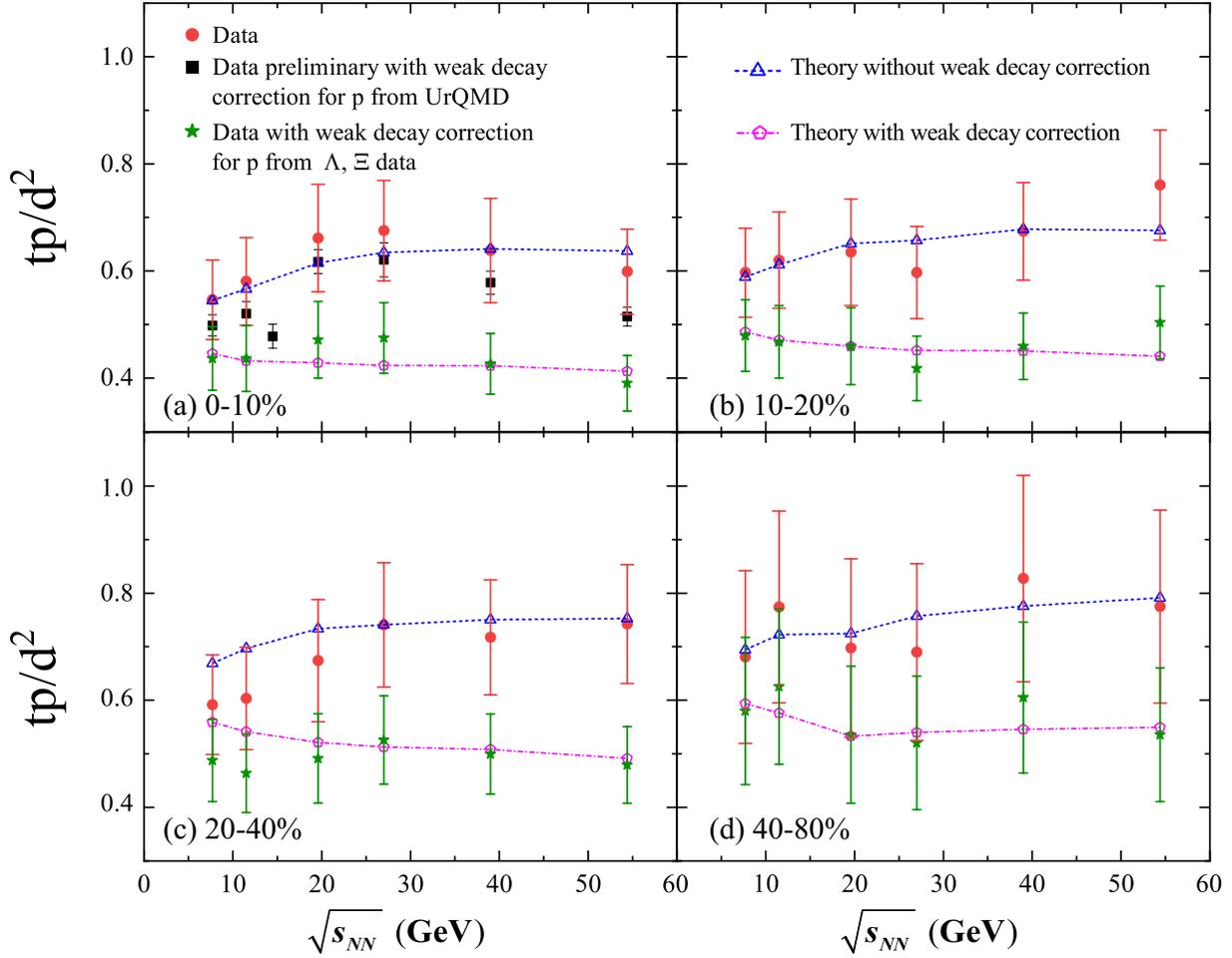}\\
	\caption{Energy dependence of $tp/d^2$ in different centralities at midrapidity in Au-Au collisions. 
Data denote by filled circles are obtained based on the yields of tritons, deuterons and final-state protons in Refs.~\cite{STAR:2017sal,B2d2019PRCSTAR,DWZhang2021NPASTAR}, and data preliminary with weak decay correction for $p$ from UrQMD denoted by filled squares are from Ref.~\cite{DWZhang2021NPASTAR}. 
Theoretical results without weak decay corrections for protons are open triangles and those with weak decay corrections are open pentagons.}
	\label{fig:tpd2-010}
\end{figure*}

In this subsection, we study the multi-particle yield correlation $tp/d^2$, which has recently attracted extensive attention~\cite{Sun:2017xrx,Sun:2020pjz,Sun:2020uoj,Sun:2020zxy,ZhaoWB2021PLB,ZhaoWB2020PRC} and considered to be a probe for the structure of the QCD phase diagram~\cite{PLB805LiuHui}. 
Compared with other yield ratios discussed in the last subsection, it has been observed by the STAR experiment to show a non-monotonic trend as the function of $\sqrt{s_{NN}}$ in the most central Au-Au collisions as shown by the solid circles and squares with error bars in Fig.~\ref{fig:tpd2-010} (a). 
Note that solid circles are obtained based on the yields of tritons, deuterons and final-state protons measured in Refs.~\cite{STAR:2017sal,B2d2019PRCSTAR,DWZhang2021NPASTAR}, 
and solid squares are the STAR preliminary data in which the proton yield has been corrected by weak-decay feeddown from strange baryons based on the UrQMD simulation ~\cite{STAR:2018PRC}.
The peak around 20 GeV in these solid symbols is considered in some works to be a signal of an enhanced baryon density fluctuation and therefore a possible signal of 
potentially a critical point~\cite{PLB805LiuHui,KJS2021EPJA}.

To further ascertain the reliability of $tp/d^2$ as a probe of a large baryon density fluctuation near the critical point and/or production mechanism of light nuclei, different methods of correcting the weak decay contamination for protons are necessary.
We here use a data-driven weak decay correction for the proton in $tp/d^2$, i.e., $p_{\text{withWDC}} \approx p-63.9\%\Lambda$, in which $p$ and $\Lambda$ denote the inclusively measured proton and $\Lambda$ at experiment.
The filled stars with error bars in Fig.~\ref{fig:tpd2-010} (a) are the experimental data after correcting $\Lambda$ and $\Xi$ weak decay contaminations~\cite{STAR:2019bjj} for protons, which are much smaller than those solid circles and also much smaller than those solid squares.
Peak behavior around 20 GeV is weakened in the $\Lambda$ and $\Xi$ data-driven weak decay correction result for $tp/d^2$ compared to that without weak decay corrections.
Open triangles connected with lines to guide the eye are our results without weak decay corrections, which basically agree with the data and exhibit an increasing trend and then seems invariant at $\sqrt{s_{NN}}>27$ GeV.
Open pentagons connected with lines to guide the eye are our corresponding theoretical results with weak decay corrections, where we use SDQCM to correct the decay contaminations from strange hyperons.
They exhibit very slightly decreasing trend and agree with the data within error bars.
Compared the result without weak decay correction with that with weak decay correction, one can see that protons from hyperon weak decays have different influences on the behavior of $tp/d^2$ at different collision energies, i.e., 
the weak decay contamination for $tp/d^2$ is different at different collision energies.

We also study $tp/d^2$ in other centralities and results are given in Fig.~\ref{fig:tpd2-010} (b), (c), (d).
In these three centralities, there seems no peak behaviors as the function of $\sqrt{s_{NN}}$.
Theoretical results without weak decay corrections denoted by open triangles increase slightly and then become invariant,
and those with weak decay corrections denoted by open pentagons decrease slightly.
All theoretical results agree with the data within error bars.
The different behaviors for open triangles and pentagons come from different contributions of protons from hyperon weak decays.
With the increasing energy, decay contribution becomes stronger.
This makes the decreasing trend of without-weak-decay-correction results to be increasing or invariant behavior of with-weak-decay-correction results.
This further show that $tp/d^2$ should be carefully corrected from hyperon weak decays for protons to probe the production characteristics of light nuclei and the structure of the QCD phase diagram.

\section{summary}

Based on the description of the production of different mesons and baryons by the SDQCM, we obtained $p_T$ distributions of final-state (anti-)protons as well as those produced at the kinetic freeze-out in Au-Au collisions at $\sqrt{s_{NN}}=$7.7, 11.5, 19.6, 27, 39, 54.4 GeV.
We found that weak decay contaminations for protons from $\Lambda$ and $\Sigma$ hyperons were different in different centralities at different collision energies.
With these momentum distributions of kinetic freeze-out protons and antiprotons obtained from the SDQCM, we studied the production of light nuclei and antinuclei in the (anti-)nucleon coalescence mechanism in relativistic heavy ion collisions at energies employed at the RHIC beam energy scan.

We firstly computed the $p_T$ spectra of (anti-)deuterons ($d$, $\bar d$) and (anti-)tritons ($t$, $\bar t$) in Au-Au collisions at $\sqrt{s_{NN}}=$7.7, 11.5, 19.6, 27, 39, 54.4 GeV
and found the available experimental data for these $p_T$ spectra can be well reproduced.
On this basis, we gave the predictions of $p_T$ spectra for $\bar d$ at $11.5$ GeV for $60-80\%$ and $7.7$ GeV and $\bar t$ at these energies.
We then studied the yield densities of light (anti-)nuclei and our results were consistent with the available data.

We finally studied different yield ratios, such as $\bar d/d$,  $\bar t/t$,  $d/p$, $\bar d/\bar p$,  $t/p$, $\bar t/\bar p$,  $t/d$, $\bar t/\bar d$, $d/p^2$, $\bar d/\bar p^2$, etc.,  and naturally explained their interesting behaviors as the function of the collision energy.
We especially studied the multi-particle yield correlation $tp/d^2$ and pointed out that it should be carefully corrected from hyperon weak decays for protons to employ it to probe the production characteristics of light nuclei and the structure of the QCD phase diagram.
All of our results showed that the coalescence mechanism for nucleons and antinucleons played a dominant role for the production of light nuclei and antinuclei at RHIC beam energy scan energies.

\section*{Acknowledgements}

We thank Xiao-Feng Luo for helpful discussions.
This work was supported in part by the National Natural Science Foundation of China under Grant No. 12175115 and No. 11975011, the Natural Science Foundation of Shandong Province, China, under Grants No. ZR2020MA097, No. ZR2019YQ06 and No. ZR2019MA053, and Higher Educational Youth Innovation Science and Technology Program of Shandong Province under Grants No. 2020KJJ004 and No. 2019KJJ010.

\end{document}